\documentclass[aps,twocolumn,reprint,groupedaddress,floatfix]{revtex4}

\usepackage{graphicx}
\usepackage{amsmath,amssymb}
\usepackage{mathtools}
\usepackage{bm}
\usepackage{color}
\usepackage[normalem]{ulem}
\usepackage[usenames,dvipsnames]{xcolor}
\usepackage{eufrak}

\newcommand{\Hvec}{\mathbf{H}}

\newcommand{\Qvec}{\mathbf{Q}}
\newcommand{\Qij}{Q_{ij}}

\DeclareMathOperator{\Tr}{Tr}

\begin{document}

\title{{Nematic Films at Chemically Structured Surfaces}}

\author{N. M. Silvestre}
\author{M. M. Telo da Gama}
\affiliation{Centro de F\'\i sica Te\'orica e Computacional, Departamento de F\'\i sica, Faculdade 
de Ci\^encias, Universidade de Lisboa, Campo Grande P-1749-016 Lisboa, Portugal}
\author{M. Tasinkevych}
\email[Corresponding author: ]{miko@is.mpg.de}
\affiliation{Max-Planck-Institut f\"{u}r Intelligente Systeme, Heisenbergstr. 3, 
D-70569 Stuttgart, Germany}
\affiliation{IV. Institut f\"ur Theoretische Physik, Universit{\"a}t Stuttgart, 
Pfaffenwaldring 57, D-70569 Stuttgart, Germany}

\date{\today}

\begin{abstract}
\noindent We investigate theoretically the morphology of a thin nematic film adsorbed at flat substrate patterned by stripes with alternating aligning properties, normal and tangential respectively. We construct a simple ``exactly-solvable'' effective interfacial model where the liquid crystal distortions are accounted for via an effective interface potential. We find that chemically patterned substrates can strongly deform the nematic-air interface. The amplitude of this substrate-induced undulations increases with decreasing average film thickness and with increasing surface pattern pitch. We find a regime where the interfacial deformation may be described in terms of a material-independent universal scaling function. Surprisingly, the predictions of the effective interfacial model agree semi-quantitatively with the results of the numerical solution of a full model based on the Landau-de Gennes theory coupled to a square-gradient phase field free energy functional for a two phase system. 
\end{abstract}

\maketitle

\section{Introduction}

The reduction in length scales brought about by ``lab-on-a-chip'' applications has raised a number of challenging questions. One is how to engineer devices for manipulating fluids at the micro- and nanometer length scales. The behaviour of fluids at the micron scale and below is determined mainly by viscous and surface forces, with inertia and gravity playing a negligible role. As a consequence, the structure and properties of the confining surfaces may be exploited to efficiently manipulate the behaviour of fluids which are in contact with them. Recent progress in controlled fabrication of patterned solid surfaces at the nano- to micro-meter range \cite{xia.1998,delCampo:2008,Hofmann2010,Aizenberg:2012,McConney:2013,Checo:2014} open excellent prospects in tailoring interfacial forces which in turn can be used to control the behaviour of adsorbed fluids.
Apart from an academic interest, a fundamental understanding of such systems will pave the way for the design and manufacturing of novel  materials, devices, and applications. Examples include smart surfaces with self-cleaning \cite{Tuteja:2007}, enhanced heat transfer \cite{Miljkovic:2012}, reduced fluid drag \cite{Bocquet:2011} or photo-responsive properties \cite{martinez:2011pnas}; template-directed assembly of colloidal particles \cite{Blaaderen.1997a, Rycenga.2009,silvestre:2014}; microfluidics \cite{Stone:2004,Squires:2005} and optofluidics \cite{Psaltis:2006,Monat:2007,xu:2013}. 

From the viewpoint of optical applications liquid crystals (LCs), due to their unique optical properties \cite{Tocnaye:2004,Khoo:2009}, continue to play a major role in the display industry \cite{Kim:2009}. In addition, LCs are the most representative class of room temperature ordered fluids.  As such they provide an ideal testing ground for fundamental theories of order-disorder transitions and the role of topological defects, which are of crucial importance in all areas of physics as well as in many branches of materials science. One of the most notable features of LCs is their ability to generate complex profiles of the director field at the mesoscopic level in the presence of surfaces. The surface structure, e.~g. topographical or chemical patterns, may cause frustration in the LC molecular orientation field, leading to the emergence of topological defects, which in turn may affect the nematic ordering at the surface. This provides a tunable mechanism for designing switchable surfaces for LC displays since the behaviour of defects, as the surface response to the presence of the liquid crystal, may be tuned by temperature and surface topography or chemistry. 

Nematic configurations induced by a single substrate patterned with stripes of mixed perpendicular and parallel anchoring 
properties have been studied theoretically in refs.~\cite{Qian:1997,Kondrat:2001,Atherton:2006,Harnau:2005} and  experimentally in refs.~\cite{Oo:2006,Lee:2001,Bramble:2007}. Theoretical analysis predicted the existence of both orientationally uniform and non-uniform nematic textures in a semi-infinite geometry depending on the pattern geometry and anchoring strengths.
When the nematic LC is confined between two substrates, with one micro-patterned, it is possible to create non-uniform LC textures in the cell \cite{Chen:1995,Bramble:2007}. Such micro-textures consist of alternating uniform and hybrid aligned nematic configurations (planar at one and perpendicular at the other substrate) and are used as low cost, easy-to-fabricate LC diffraction gratings \cite{Chen:1995} -- optical devices operating in a transmission mode. Similar effects may be achieved by using an array of interdigitated electrodes beneath one of the confining substrates \cite{Brown:2002}. 

In all the work mentioned above, the LC was confined in cells bounded by solid substrates. Additional exciting opportunities emerge when fluid-fluid interfaces are used as active optical surfaces, e.~g. by replacing one of the confining solid substrates. This adds a high degree of functional adaptability to optical devices as in the case of tunable-focus liquid lenses \cite{Mishra:2016}, electrowetting displays \cite{hayes:2003,Zhou:2009}, or responsive diffraction gratings \cite{brown:2009,Brown:2010,wells:2011,hsieh:2015}. 
In refs.~\cite{brown:2009,Brown:2010} dielectrophoresis  induced interfacial modulations of a free oil interface has been reported. A micron-sized oil film was deposited on a planar substrate patterned with an interdigitated array of electrodes. The oil-air interfacial deformations were achieved with the application of a voltage $\cal{V}$ difference between adjacent electrodes. The peak-to-peak amplitude $A$ of the interfacial deformations was found to behave $\sim {\cal V}^2$, and relatively high voltages  $\sim 100$ V were needed to obtain $A$ on the order of the micrometer, for average film thicknesses $\sim 10 \,\mu$m.  Here we demonstrate that similar effects may be observed, in the absence of external fields, when instead of isotropic liquids orientationally ordered nematic LCs are used as the wetting fluid. When a thin film of LC  is spread over a chemically patterned surface, the interfacial deformations emerge spontaneously as a result of the interplay between the LC elasticity and capillarity.

\begin{figure}[]
\includegraphics[width=0.95\linewidth]{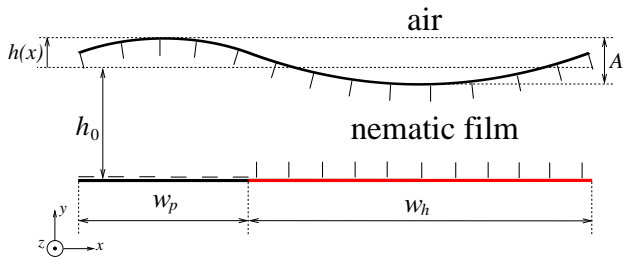}
\caption{\label{fig:geometry} 
Cross-sectional illustration of the system geometry: a nematic film of thickness $h_0 +h(x)$ is placed on top
of a chemically inhomogeneous surface. The surface consists of "planar`` stripes of width $w_p$, i.e.,  they impose
planar anchoring of the nematic director $\bf n$, and "homeotropic'' stripes of width $w_h$, i.e., they induce homeotropic (perpendicular) anchoring of $\bf n$. The nematic-air interface imposes homeotropic anchoring of $\bf n$. 
The volume per unit length $V= h_0 (w_p+w_h)$ of the nematic liquid is assumed constant. The system is translationally invariant in $z$-direction.
}
\end{figure}

In the following we investigate the modulation of nematic wetting films adsorbed on substrates decorated with stripes of alternating perpendicular and tangential anchoring properties (see Fig.~\ref{fig:geometry}). The main objective is to relate the morphology of the nematic-air interface to the average film thickness and the symmetry of the substrate pattern. In the next section we introduce a simple, ``exactly solvable''  effective interfacial model, which predicts a simple scaling behaviour of the interfacial deformations. Then, in section~\ref{sec:LDG_model} we compare the analytical predictions of the interfacial model with the numerical results of a model which fully accounts for the nematic degrees of freedom.

\section{Effective interfacial model}
\label{sec:ef_model}

We consider a nematic film of average thickness $h_0$, and with constant volume $V$ per unit length, in contact with a planar substrate chemically patterned with a one-dimensional array of alternating ``planar'' and ``homeotropic'' stripes, which impose parallel and perpendicular
anchoring boundary conditions on the nematic director field ${\bf n}({\bf r})$, respectively, see Fig.~\ref{fig:geometry}. The ``upper'' surface of the film is in contact with air where strong homeotropic anchoring conditions are assumed. The presence of the 
stripes with strong planar anchoring generates a director profile ${\bf n}({\bf r})$ which is characterised by strong distortions in the $y-$direction. These,
combined with the condition of fixed nematic volume give rise to deformations of the nematic-air interface $h(x)$, relative to $h_0$, as shown in Fig.~\ref{fig:geometry} schematically. The aim of the present work is to determine $h(x)$ as a function of various model parameters such as $h_0$, the widths $w_p, w_h$ of the planar and homeotropic stripes, the nematic elastic constants, and the surface tension of the nematic-air interface. 

The amplitude $A$ of the interfacial deformations $h(x)$ is determined (roughly) by the relative weight
of the capillary and elastic energies which depend upon $h(x)$ and ${\bf n}({\bf r})$. It is a challenging 
task to determine the nematic director field for a domain with a ``moving'' boundary, which requires lengthy numerical calculations. In order to gain insight into the behaviour of the system, we start by constructing an effective interfacial model which depends on a $1D$ (the system is translationally invariant in the $z$-direction) interfacial field $h(x)$ and where the ``bulk'' degrees of freedom ${\bf n}({\bf r})$ are ``integrated out''. In the effective model the elastic energy stored
in the director distortions is accounted for through an effective interface potential $W(x,h;h_0)$. The effective model is expressed in terms of a $1D$ free energy functional 
\begin{equation}
 {\cal F}_{eff} = \int_{0}^{w} \Bigr (\frac{\gamma}{2}(\partial_x h)^2 + W(x,h;h_0) + \lambda h\Bigl )dx,
\label{eq:eff_1D_model}
\end{equation}
where $\gamma$ is the surface tension of the nematic-air interface, and the last term with the Lagrange multiplier $\lambda$ ensures that the nematic volume is constant. The integral is taken over a unit cell of the surface pattern, with width $w = w_p+w_h$. The equilibrium shape of the nematic-air interface minimizes the functional Eq.~(\ref{eq:eff_1D_model}).

\subsection{Effective interface potential}

In order to obtain an expression for the effective interface potential $W$, we 
restrict ${\bf n}({\bf r})$ to the $(x,y)$ plane, which allows the nematic director to be written as a scalar angular orientation filed $\theta(x,y)$, 
${\bf n}=(\cos\theta,\sin\theta,0)$. Then, adopting the one elastic constant approximation,
the Frank-Oseen free energy functional (per unit length) of the bulk nematic phase is written as
 \begin{eqnarray}
{\cal F}_{el} =\frac{K}{2} \int (\nabla\theta)^2dS,
\label{eq:frank_model}
\end{eqnarray}
where $K$ is the splay-bend elastic constant in the one-constant approximation, and the integral is taken over the $(x,y)$ cross-section of the nematic film. Boundary terms are ignored in Eq.~(\ref{eq:frank_model}) which is justified in the strong anchoring regime (fixed orientation at the boundaries). For a nematic film under hybrid anchoring conditions (e.g., planar at the bottom surface $y = 0$ and homeotropic at the top $y = \hat{h}$) and with uniform thickness $\hat{h}$, the orientation field $\theta$ may be approximated by $\theta(y) =\pi y/(2 \hat{h})$. This gives
a thickness dependent elastic energy per unit area $f_{el} = K\pi^2/(8\hat{h})$ which is stored in the hybrid nematic configuration. A similar nematic film, confined by homeotropic surfaces, stores no elastic energy. Based upon these observations we approximate the effective interface potential in Eq.~(\ref{eq:eff_1D_model}) as follows:

\begin{equation}
    W(x,h;h_0) = 
    \begin{cases}
      \frac{K\pi^2}{8(h_0+h)} &  0\leqslant x \leqslant w_p,\\
       0        &  w_p< x \leqslant w,
    \end{cases}
\label{eq:eff_pot}
  \end{equation}
 where we have taken $\hat{h}=h_0+h(x)$.

\subsection{The interfacial deformation $h(x)$}

\begin{figure*}[th]
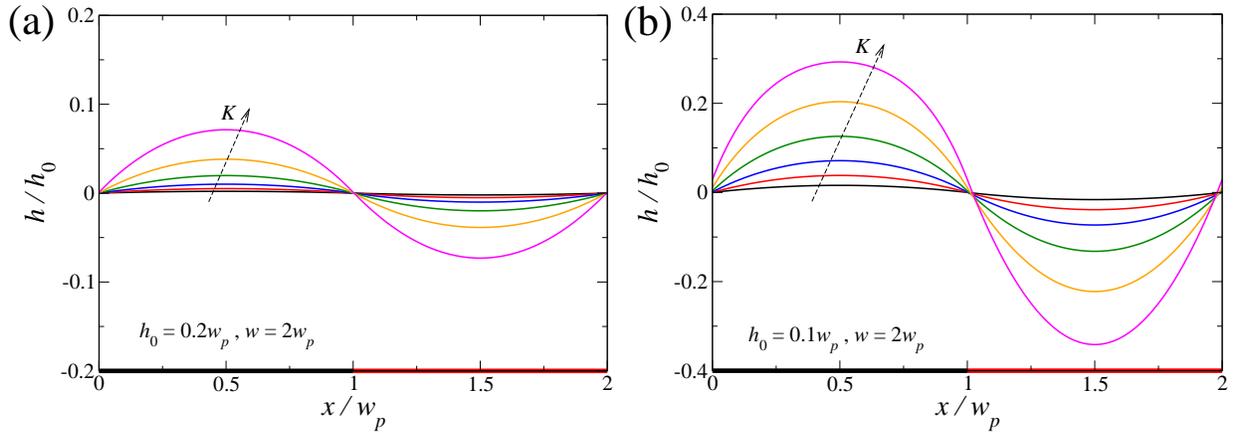

\includegraphics[width=0.45\linewidth]{fig02a.eps}
\includegraphics[width=0.45\linewidth]{fig02b.eps} \hspace*{0.5cm}
\caption{\label{fig:profiles_vs_K} 
A film of non-volatile nematic liquid at a planar substrate patterned with a one-dimensional array of stripes with planar and homeotropic anchoring. The width of the planar stripes $w_p$ is equal to the width of the homeotropic stripes $w_h$. Substrate-induced deformation $h(x)$ of the nematic-air interface measured with respect to a reference height $h_0$ (see 
main text for details) for several values of the the splay-bend elastic constant $K$.   (a) $h_0=0.2w_p$; (b) $h_0 =0.1w_p$. Different curves correspond to  $K/(\gamma h_0^{(a)}) \approx  0.001, 0.003, 0.005, 0.01, 0.02, 0.04$. $w_p=w_h=117.5 K_{max}/\gamma$ are kept fixed.
}
\end{figure*}

\begin{figure*}[th]
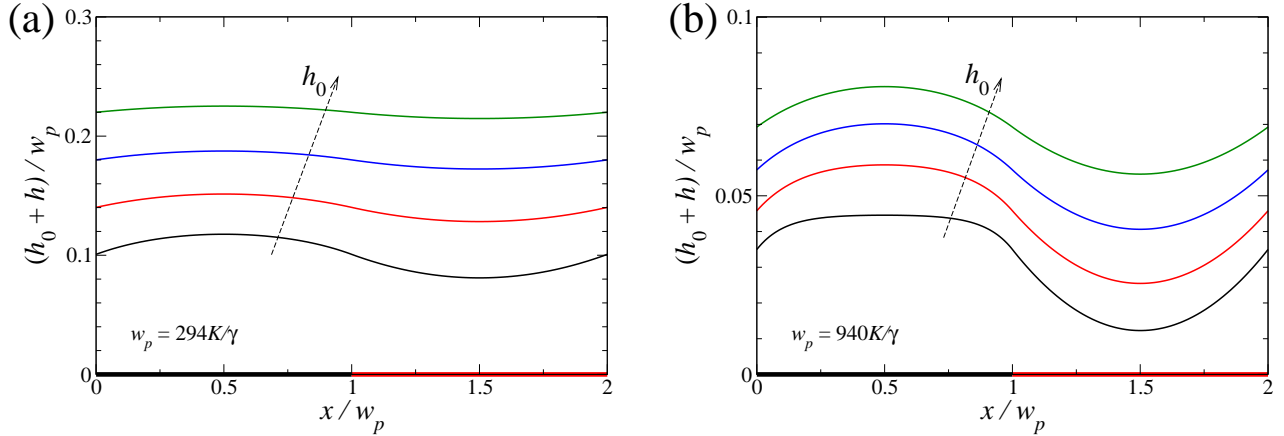

\includegraphics[width=0.45\linewidth]{fig03a.eps} \hspace*{0.5cm}
\includegraphics[width=0.45\linewidth]{fig03b.eps}
\caption{\label{fig:profiles_vs_h0} 
A film of non-volatile nematic liquid at a planar substrate patterned with a one-dimensional array of stripes with planar (width $w_p$) and homeotropic (width $w_h$) anchoring; $w_p = w_h$. Substrate-induced deformation $h(x)$ of the nematic-air interface relative to a reference height $h_0$ (see main text for details) for several values $h_0$. The curves in (a) and (b) with the same color correspond 
to the same value of $h_0$, but different $w_p$ with the ratio $w_p^{(b)}/w_p^{(a)}=3.2$. The curves from bottom to top correspond to $h_0 \gamma / K\approx 29,  41, 53, 65$.}
\end{figure*}

Minimizing  Eq.~(\ref{eq:eff_1D_model}) with respect to $h$ and $\lambda$ leads to the corresponding Euler-Lagrange equation
\begin{equation}
 \gamma \partial_x^2 h- \partial_h W(x,h;h_0) -\lambda =0,
\label{eq:Euler-Lagrange}
\end{equation}
and the constraint of constant volume 
\begin{equation}
 \int_{0}^{w} h(x)dx = 0.
\label{eq:constant_volume}
\end{equation}
We are interested in the solution to Eq.~(\ref{eq:Euler-Lagrange}) for $x\in[0,w]$ with periodic boundary conditions, i.e.,
$h(0)=h(w)$, and $\partial_x h(x)|_{x=0}=\partial_x h(x)|_{x=w}$. When $h(x)\ll h_0$, Eq.~(\ref{eq:Euler-Lagrange}) may be linearized and its solution may be written in the form
\begin{equation}
    h(x) = 
    \begin{cases}
           h_p(x) &  0\leqslant x \leqslant w_p,\\
      h_h(x)  &  w_p< x \leqslant w.
    \end{cases}
\label{eq:solution_general}
  \end{equation}
 It is easy to find that 
\begin{eqnarray}
 h_p(x) &=& \frac{h_0}{2} - \lambda\frac{4h_0^3}{\pi^2 K} + c_1\exp(-q x) + c_2\exp(q x),
\label{eq:solution1}
 \\
h_h(x) &=& c_3 + c_4 x + \frac{\lambda x^2}{2 \gamma},
\label{eq:solution2}
\end{eqnarray}
where $q = \frac{\pi}{2h_0}\sqrt{\frac{K}{h_0\gamma}}$. The unknown integration constants $c_i, i=1,..,4$ and the Lagrange multiplier $\lambda$ are determined from the periodic boundary conditions, the condition that $h(x)$ must be smooth at $x = w_p$, and the constant volume constraint Eq.~(\ref{eq:constant_volume}). Throughout this work we keep $\gamma$ fixed. Most of the results,
presented below, except the two last figures,  are obtained for symmetric chemical patterns with $w_p = w_h$.

Since the interfacial deformation is driven by the ``repulsive'' effective interface potential $W\propto K/(h_0+h)$ it is natural to expect that $h(x)$ and the peak-to-peak amplitude $A$ will grow with growing $K$ and decreasing average film thickness $h_0$. In Fig.~\ref{fig:profiles_vs_K}  we plot $h(x)$ given by Eqs.~(\ref{eq:solution1}), (\ref{eq:solution2}) for several values of the bend-splay elastic constant $K$. $h_0$ in Fig.~\ref{fig:profiles_vs_K}(a) is twice of that in Fig.~\ref{fig:profiles_vs_K}(b).  As expected the amplitude of the interfacial deformations grows with $K$. Thus, at large $h_0$ we find $A/h_0 \simeq 0.004$ and  $\simeq 0.14$ for the smallest and the largest $K$, respectively; while at small $h_0$ the ratios are $A/h_0 \simeq 0.03$ and $\simeq 0.63$, respectively. The shape of the interface is roughly antisymmetric, i.e, $h_p(x/w_p) = - h_h(x/w_p-1)$, for small to intermediate values of $K$. At the largest value of $K$ and at $h_0 = 0.1w_p$, a clear difference between  the ``left'' (above the planar stripe) part of the interface $h_p(x)$ and the ``right'' (above the homeotropic stripe)  one $h_h(x)$ is observed (magenta curve in Fig.~\ref{fig:profiles_vs_K}(b)). 

Upon further increase (decrease) of $K$ ($h_0$) this left-right asymmetry becomes more pronounced, see for example the black curve in Fig.~\ref{fig:profiles_vs_h0}(b). However, the results for strongly deformed interfaces $h(x)$ with peak-to-peak amplitudes $A$ comparable to $h_0$, should be taken with caution as the linearised solution of Eq.~(\ref{eq:eff_1D_model}) breaks down in this regime. Figure~\ref{fig:profiles_vs_h0} demonstrates in more detail the effects of varying the average film thickness $h_0$ on the shape of the interfacial deformation, at fixed $K$. Again, $A$ decreases with increasing $h_0$, as expected, and the profiles start to exhibit the antisymmetric property $h_p(x/w_p) = - h_h(x/w_p-1)$ around $x/w_p = 1$. 
In Fig.~\ref{fig:profiles_vs_h0} we illustrate the effect of increasing the pitch $w$ of the symmetric chemical pattern, which leads to an increase of $A$.

A closer inspection of the interfacial profiles obtained for different values of the model parameters and for $A/h_0\ll 1$, suggests that a scaling solution exists in this regime. Therefore, we pose the following question: what are the scaling properties of $h(x)$ as a function of $K, h_0$ and $w_p$ in the regime of small interfacial deformations? We address the question in the next section using a simple one-parametric Ansatz.

\subsection{A scaling function for $h(x)$}

\begin{figure}
\includegraphics[width=0.95\linewidth]{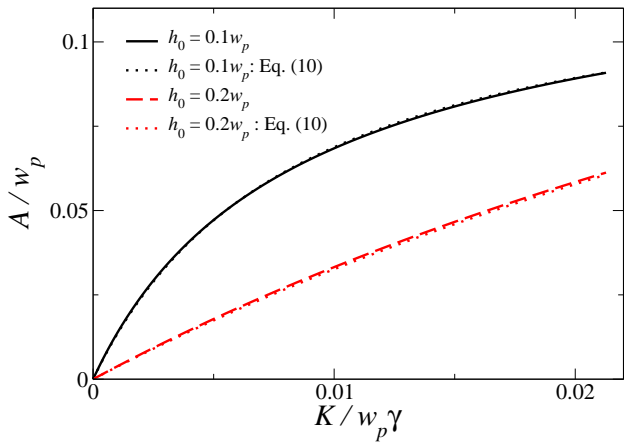}
\caption{\label{fig:delta_h_vs_K} 
Peak-to-peak amplitude $A$ of the interfacial deformation $h(x)$ as a function of the bend-splay elastic constant $K$ for two values of the average film thickness $h_0$  at fixed $w_p=w_h$. }
\end{figure}

\begin{figure}[]
\includegraphics[width=0.95\linewidth]{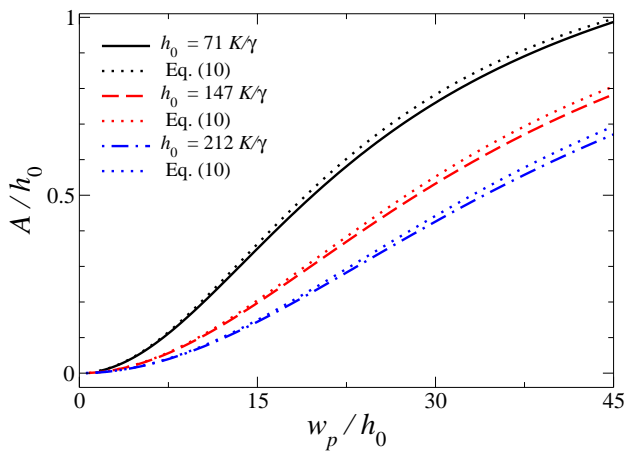} 
\caption{\label{fig:delta_h_vs_W_diffh0} 
Peak-to-peak amplitude $A$ of the interfacial deformation $h(x)$ as a function of the width of the planar stripe $w_p$ with $w_h=w_p$ for several values of the average film thickness $h_0$. 
} 
\end{figure}

\begin{figure}[]
\includegraphics[width=0.95\linewidth]{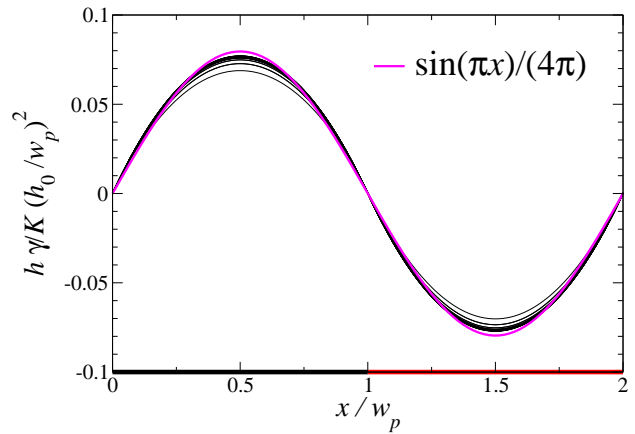} 
\caption{\label{fig:scaling_function}
Universal scaling function $\cal {H}$, defined in Eq.~(\ref{eq:scaling_forms}) in the regime $qw_p\rightarrow 0$, of the substrate-induced interfacial deformation $h(x)$. The shape of the scaling function is similar to that given by the ansatz Eq.~(\ref{eq:ansatz}), but its amplitude is reduced by a factor $R\simeq 0.97$.
}
\end{figure}

To investigate the scaling behaviour of $h(x)$, and motivated by the simple shape of the interface for $A/h_0\ll 1$, we use a sinusoidal Ansatz 
\begin{equation}
\tilde{h}(x) = \frac{\tilde{A}}{2} \sin (\pi x/w_p ),
\label{eq:ansatz}
\end{equation}
with a single variational parameter $\tilde{A}$, which satisfies the fixed volume constraint Eq.~(\ref{eq:constant_volume}). Substituting $\tilde{h}(x)$ into the free energy functional Eq.~(\ref{eq:eff_1D_model}), within the quadratic approximation in $h$ for the effective interface potential Eq.~(\ref{eq:eff_pot}),  and minimizing the resulting function with respect to $\tilde{A}$ yields
\begin{equation}
 \tilde{A} = \frac{K}{\gamma}\biggl(\frac{w_p}{h_0}\biggr)^2\frac{\pi}{2\pi^2 + (qw_p)^2}.
\label{eq:ansatz_scalling_amplitude}
\end{equation}
This expression provides a very good estimate of the peak-to-peak interfacial amplitude $A$ as shown in Fig.~\ref{fig:delta_h_vs_K} and Fig.~\ref{fig:delta_h_vs_W_diffh0}. In the figures the analytical amplitude 
$\tilde{A}$ is plotted as dotted lines which should be compared with the results of Eqs.~(\ref{eq:solution1}), and (\ref{eq:solution2}) plotted as solid, dashed or dashed-dotted lines. Excellent agreement between the two
approaches is found over the whole range of the independent variable, which is  $K$ in Fig.~\ref{fig:delta_h_vs_K} and $w_p$ in Fig.~\ref{fig:delta_h_vs_W_diffh0}.

Additionally, equation~(\ref{eq:ansatz_scalling_amplitude}) implies that the interfacial deformation in the limit $qw_p \rightarrow 0$ has the scaling form
\begin{eqnarray}
   h(x) &=& \frac{K}{\gamma} \Bigl(\frac{w_p}{h_0}\Bigr )^2{\cal H}(x/w_p)  
  \label{eq:scaling_forms}      \nonumber  \\
&=&h_0(qw_p)^2\frac{4}{\pi^2}{\cal H}(x/w_p), 
\end{eqnarray}
where ${\cal H}$ is a ``universal'' scaling function that should not depend on the material properties or on the pitch $w$.  In Fig.~\ref{fig:scaling_function} we plot $\gamma (h_0/w_p)^2 h(x)/K$ obtained form Eqs.~(\ref{eq:solution1}),~(\ref{eq:solution2}) as a function of $x/w_p$ for various pattern pitches $w$, average heights $h_0$ and bend-splay elastic constants $K$. Provided that $qw_p<0.5$, scaling is observed as shown by the limiting master curve (thick black line), in very good agreement with the theoretical expression in Eq.~(\ref{eq:scaling_forms}). The leading to scaling correction is $\propto -(qw_p)^2/(2\pi)^3$ and the three thin black curves in Fig.~\ref{fig:scaling_function} that deviate most from the limiting master curve correspond (from the bottom to top) to  $qw_p \simeq 1.4, 1.0, 0.7$. In Fig.~\ref{fig:scaling_function} we also show as a magenta line the scaling function $\sin(\pi x)/(4 \pi)$ of the sinusoidal Ansatz Eq.~(\ref{eq:ansatz}). The shapes of the two curves are very similar and the amplitudes differ by a small amount, their ratio being $R\simeq0.97$.

Finally, we comment on the thermodynamic stability of the non-uniform solution. Considering the single-parameter family of interfacial shapes given by Eq.~(\ref{eq:ansatz}) the free energy per unit length of the deformed interface $\tilde{F}_{eff}=  \min_{\tilde{A}}\mathcal{F}_{eff}$  is compared to that of the flat film which is $K\pi^2 w_p/(16 h_0)$. 
The difference  is
\begin{equation}
 \tilde{F}_{eff} - \frac{K\pi^2 w_p}{16 h_0}= -\frac{K^2 w_p^3}{32\gamma h_0^4 + 4 h_0Kw_p^2}<0,
\label{eq:minimal_free_energy}
\end{equation}
implying that the deformed interface is always energetically favourable, even for arbitrarily large  $\gamma$. However, the amplitude of such deformation is $\sim 1/\gamma$, Eq.~(\ref{eq:ansatz_scalling_amplitude}). 

In order to check the validity of the effective interfacial model, Eq.~(\ref{eq:eff_1D_model}), in the next section we analyse
a mesoscopic model of the system based on the Landau-de Gennes free energy of nematics, coupled to a phase-field model of  the nematic-air interface. We shall see the results of this ``full'' model agree very well with the interfacial model approach. 

\section{Landau-de Gennes free energy coupled to a phase field model}
\label{sec:LDG_model}

\begin{figure*}[ht]
\includegraphics[width=0.85\linewidth]{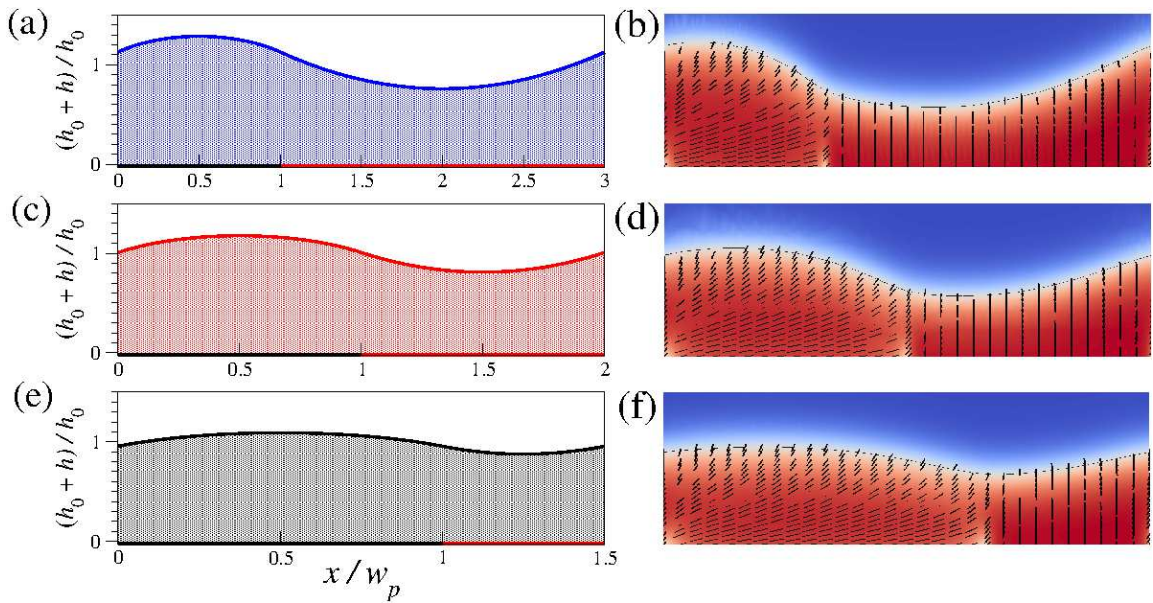}
\caption{\label{fig:profiles_numerics} 
Left: Interfacial profile $h0+h(x)$  of the nematic-air interface at a planar substrate patterned with a one-dimensional array of stripes with planar and homeotropic anchoring. The width of the planar stripe is $w_p = 294 K/\gamma$; the average film thickness $h_0 = 0.1 w_p$. The width of the homeotropic stripe a) $w_h = 2 w_p$; c)  $w_h = w_p$, and d) $w_h = 0.5 w_p$. Right: nematic configurations of the wetting films obtained by the numerical minimization of Eq.~(\ref{eq:F_ldg}). $w_p$ and $h_0$ are the same as on the left; b) $w_h = 2 w_p$; d)  $w_h = w_p$, and f) $w_h = 0.5 w_p$. Director field ${\bf n}$ is depicted by black bars. 
}
\end{figure*}

\begin{figure}[]
\includegraphics[width=0.95\linewidth]{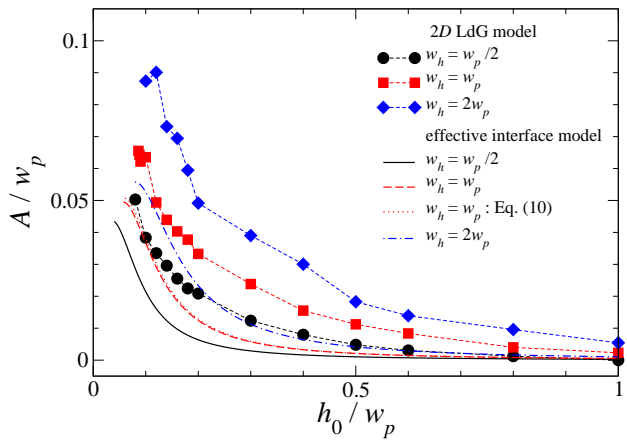}
\caption{\label{fig:delta_h_vs_h0} 
Peak-to-peak amplitude $A$ of the interfacial deformation $h(x)$ as a function of the average film thickness $h_0$ for several values of $w_h$ at $w_p = 294 K/\gamma$. (a) Results of the linear analytical theory, Eqs.~(\ref{eq:solution1}),~(\ref{eq:solution2}). (b) Numerical results obtained for the same values of parameters as in (a)  using  the Landau-de Gennes free energy coupled with a square gradient phase-field model to describe a two-phase system (see main text for details).
}
\end{figure}

We adopt a phenomenological model introduced in ref.~\cite{Takeaki:2004} to describe the dynamics of phase separation into isotropic and nematic phases. The non-dimensional free energy functional  is given by
\begin{equation}
{\cal F}_{2D}[\phi,\Qvec]=\int_\Omega{dV\left[f_{mix}(\phi)+f_{LdG}(\phi,\Qvec)+f_{int}(\phi,\Qvec)\right]}.
\label{eq:F_ldg}
\end{equation}
The free energy of mixing is given by a $\phi^4$-model:  $f_{mix}(\phi)=-\frac{1}{2}\phi^2+\frac{1}{4}\phi^4+\frac{1}{2}|\nabla\phi|^2$. $\phi=\pm 1$ are the equilibrium concentrations in the absence of nematic ordering. For simplicity the nematic liquid crystal is described by a two-dimensional tensor order parameter $\Qij=S(n_i n_j -\delta_{ij}/2)$, and contributes to the free energy as:
 $f_{LdG}=\alpha\left(-\frac{\phi}{2}\Tr{Q^2}+\frac{1}{4}\left(\Tr{Q^2}\right)^2+\frac{\varepsilon^2}{2}\left(\partial_k \Qij\right)^2\right)$. Nematic ordering proceeds as a second-order phase transition. We choose $\phi>0$ to correspond to the liquid crystal rich phase. 

The preferred nematic alignment at the interface is modelled by the coupling of $\Qij$ with the deformations of $\phi$ through: $f_{int}=\omega\,\partial_i\phi\,\Qij\,\partial_j\phi$. If $\omega>0$ the nematic will align parallel to the interface. On the other hand, if $\omega<0$ the nematic molecules will take a perpendicular alignment. 
 
 The free energy and the length are scaled by $\gamma \Xi^2$ and $\Xi$, respectively. $\gamma$ is again the interfacial tension and $\Xi$ is the correlation length associated to fluctuations of $\phi$. The additional parameters are as follows. $\alpha=K\Xi/\gamma\xi^2$ is the ratio between the elastic energy density $K/\xi^2$ and the interfacial energy density $\gamma/\Xi$, where $K$ is the Frank elastic constant defined previously and $\xi$ is the nematic correlation length. $\varepsilon=\xi/\Xi$ is  the ratio between the two correlation lengths. Typical values are: $\alpha=0.25$, $\varepsilon=0.4$ and $|\omega|=0.1$.

The evolution of $\phi$ is governed by the Cahn-Hilliard equation:
\begin{equation}
\frac{\partial}{\partial t}\phi = \nabla^2 \mu\,.
\label{eq:phi}
\end{equation}
$\mu$ is the chemical potential given by $\mu=\frac{\delta F}{\delta \phi}$.
The nematic tensor order parameter evolves through a relaxation equation:
\begin{equation}
\frac{\partial}{\partial t}\Qij =\Gamma H_{ij},
\label{eq:Q}
\end{equation}
where $\Hvec=-\left(\frac{\delta F}{\delta \Qvec}-\frac{1}{2}\mathbf{I}\Tr\frac{\delta F}{\delta \Qvec}\right)$ is the molecular field tensor.
To describe the anchoring at the chemically structured surface we consider the Nobili-Durand surface free energy density: $\frac{1}{2}\omega_s\Tr\left(\Qvec-\Qvec_s\right)^2$.

Equations (\ref{eq:phi}) and (\ref{eq:Q}) are solved with the commercial software COMSOL 3.5a \cite{comsol}, which uses finite elements techniques. We consider that the system translationally invariant and thus solved the numerics in two dimensions.
The simulation box is rectangular with width $w=w_p+w_h$, set by the width of the strips, and height $\mathrm H$. The interface is set at a height $h_0< {\mathrm H}$. Periodic boundary conditions are assumed on the lateral boundaries. As initial conditions we assume in the region close to the substrate (below $h_0$) a uniformly aligned nematic ($\phi=1$), and above the interface an isotropic fluid ($\phi=-1$). Typically we use ${\mathrm H}=5h_0$.

In Fig.~\ref{fig:profiles_numerics} we show the configurations of the nematic films as predicted by numerical minimization of the free energy functional Eq.~(\ref{eq:F_ldg}). We note that in all three cases the director profile ${\bf n}({\bf r})$ exhibit rather sharp ``interface'' between the hybrid aligned nematic and the homeotropic textures.  This could be one of the reasons why the interfacial profiles (right column in Fig.~\ref{fig:profiles_numerics}) of the full 2$D$ model are very similar to those obtained from the effective 1$D$ interfacial description (left column in Fig.~(\ref{fig:profiles_numerics}).  Fig.~\ref{fig:delta_h_vs_h0} compares the prediction of two approaches for the dependence of the peak-to-peak amplitude $A$ on the average interfacial thickness $h_0$. Here, in addition to symmetric patterns, $w_p =w_h$ (red curves and red squares), we also show results for non-symmetric patterns, with $w_h=0.5 w_p$ (black curve and black circles) and with  $w_h=2 w_p$ (blue curves and blue diamonds). We observe rather good agreement in all cases. The results based on the full model in Eq.~(\ref{eq:F_ldg}) lie systematically above the predictions of the interfacial model Eq.~(\ref{eq:eff_1D_model}). The red dotted line in Fig.~\ref{fig:delta_h_vs_h0} corresponds to the analytical expression Eq.~(\ref{eq:ansatz_scalling_amplitude}), which again agrees very well with the results of Eqs.~(\ref{eq:solution1}),~(\ref{eq:solution2}).

\section{Discussion and Conclusions}

Equation ~(\ref{eq:ansatz_scalling_amplitude}) predicts that the peak-to-peak amplitude $A$ is proportional to the intrinsic length scale $l_0=K/\gamma$ of the interfacial model Eq.~(\ref{eq:eff_1D_model}). We obtain $K/\gamma\approx 2$\AA~using $K=7$pN and
$\gamma = 35$mN/m, respectively the average elastic constant \cite{Madhusudana:1982} and the nematic-air surface tension \cite{Tintaru:2001} of 4-n-pentyl-4′-cyanobiphenyl (5CB). The corresponding $l_0$ is very small, meaning that the elasticity 
of 5CB is too small in order ``to beat '' capillarity, and extremely thin wetting films or chemical patterns with very large pitch 
$w$ must be used to produce ``measurable'' interfacial deformations. Thus, in order to obtain peak-to-peak amplitudes $A$ on the order of the micrometer one needs to use chemical patterns with $w\sim 1$mm assuming that $h_0 = 10\mu$m. 
In order to bring down the value of $w$, one must increase  $l_0$. For example, by decreasing $\gamma$ (e.g., by using surfactants, or replacing air with another fluid) and increasing the average elastic constant $K$ by a factor of $10$ respectively, the peak-to-peak amplitude is $A\sim 1\mu$m for $w\sim 100\mu$m, and $h_0 = 10\mu$m. Another possibility is to exploit the $\sim 1/h_0^2$ dependence of the amplitude on the average film thickness and to use thinner wetting films.

In summary, we investigated how chemically patterned surfaces may be used to create distortions in thin nematic wetting films and how the surface-induced distortions propagate to the free nematic-air interface. We have related the morphology of the nematic-air interface to the averaged film thickness and the parameters of the surface patterns. We have constructed a simple effective interfacial model, where the nematic degrees of freedom are ``integrated out''. This model predicts 
a simple scaling behaviour of the interfacial deformation profile $h(x)$, which is characterised by a material-independent universal scaling function. Our results provide new insights into the effects of the coupling of the orientational elasticity and capillarity and may be used to design responsive LC-based diffraction gratings which could operate in both transmission and reflection modes.

\acknowledgments 
We acknowledge financial support of the Portuguese Foundation for Science and Technology (FCT) under the contracts numbers UID/FIS/00618/2013 and EXCL/FIS-NAN/0083/2012. 

\providecommand{\noopsort}[1]{}\providecommand{\singleletter}[1]{#1}%

\end{document}